\documentclass[12pt]{article}
\usepackage{latexsym,amssymb,amsmath}
\begin{document}
\baselineskip=20pt

\newcommand{\la}{\langle}
\newcommand{\ra}{\rangle}
\newcommand{\psp}{\vspace{0.4cm}}
\newcommand{\pse}{\vspace{0.2cm}}
\newcommand{\ptl}{\partial}
\newcommand{\dlt}{\delta}
\newcommand{\sgm}{\sigma}
\newcommand{\al}{\alpha}
\newcommand{\be}{\beta}
\newcommand{\G}{\Gamma}
\newcommand{\gm}{\gamma}
\newcommand{\vs}{\varsigma}
\newcommand{\Lmd}{\Lambda}
\newcommand{\lmd}{\lambda}
\newcommand{\td}{\tilde}
\newcommand{\vf}{\varphi}
\newcommand{\yt}{Y^{\nu}}
\newcommand{\wt}{\mbox{wt}\:}
\newcommand{\rd}{\mbox{Res}}
\newcommand{\ad}{\mbox{ad}}
\newcommand{\stl}{\stackrel}
\newcommand{\ol}{\overline}
\newcommand{\ul}{\underline}
\newcommand{\es}{\epsilon}
\newcommand{\dmd}{\diamond}
\newcommand{\clt}{\clubsuit}
\newcommand{\vt}{\vartheta}
\newcommand{\ves}{\varepsilon}
\newcommand{\dg}{\dagger}
\newcommand{\tr}{\mbox{Tr}}
\newcommand{\ga}{{\cal G}({\cal A})}
\newcommand{\hga}{\hat{\cal G}({\cal A})}
\newcommand{\Edo}{\mbox{End}\:}
\newcommand{\for}{\mbox{for}}
\newcommand{\kn}{\mbox{ker}}
\newcommand{\Dlt}{\Delta}
\newcommand{\rad}{\mbox{Rad}}
\newcommand{\rta}{\rightarrow}
\newcommand{\mbb}{\mathbb}
\newcommand{\lra}{\Longrightarrow}
\newcommand{\X}{{\cal X}}
\newcommand{\Y}{{\cal Y}}
\newcommand{\Z}{{\cal Z}}
\newcommand{\U}{{\cal U}}
\newcommand{\V}{{\cal V}}
\newcommand{\W}{{\cal W}}
\newcommand{\sech}{\mbox{sech}\:}
\newcommand{\csch}{\mbox{csch}\:}
\newcommand{\sn}{\mbox{sn}\:}
\newcommand{\cn}{\mbox{cn}\:}
\newcommand{\dn}{\mbox{dn}\:}

\begin{center}{\Large \bf Quadratic-Argument Approach to Nonlinear}\end{center}
\begin{center}{\Large \bf   Schr\"{o}dinger Equation and Coupled Ones}
\end{center}
\vspace{0.2cm}

\begin{center}{\large Xiaoping Xu}\end{center}
\begin{center}{Institute of Mathematics, Academy of Mathematics \& System Sciences}\end{center}
\begin{center}{Chinese Academy of Sciences, Beijing 100190, P.R. China}
\footnote{Research supported
 by China NSF 10431040}\end{center}

\vspace{0.6cm}

 \begin{center}{\Large\bf Abstract}\end{center}

\vspace{1cm} {\small The two-dimensional cubic nonlinear
Schr\"{o}dinger equation is used to describe the propagation of an
intense laser beam through a medium with Kerr nonlinearity. The
coupled two-dimensional cubic nonlinear  Schr\"{o}dinger equations
 are used to describe interaction of electromagnetic
waves with different polarizations in nonlinear optics.  In this
paper, we solve the above equations  by imposing a quadratic
condition on the related argument functions and using their symmetry
transformations. More complete families of exact solutions of such
type are obtained. Many known interesting solutions, such soliton
ones, turn out to be special cases of our solutions.}

\section{Introduction}

The two-dimensional cubic nonlinear  Schr\"{o}dinger equation:
$$ i\psi_t+c(\psi_{xx}+\psi_{yy})+a|\psi|^2\psi=0\eqno(1.1)$$
is used to describe the propagation of an intense laser beam
through a medium with Kerr nonlinearity, where $t$ is the distance
in the direction of propagation, $x$ and $y$ are the transverse
spacial coordinates,  $\psi$ is a complex valued function in
$t,x,y$ standing for electric field amplitude, and $a,c$ are
nonzero real constants. We refer the introduction of [SEG] for
more systematic exposition of the equation. Akhnediev, Eleonskii
and Kulagin [AEK] found certain exact solutions of (1.1) whose
real and imaginary parts are linearly dependent over the functions
of $t$. Moreover, Gagnon and Winternitz [GW] found exact solutions
of the cubic and quintic nonlinear Schr\"{o}dinger equation for a
cylindrical geometry. Mihalache and Panoin [MN] used the method in
[AEK] to obtain new solutions which describe the propagation of
dark envelope soliton light pulses in optical fibers in the normal
group velocity dispersion regime. Furthermore, Saied, EI-Rahman
and Ghonamy [SEG] used various similarity variables to reduce the
above equation to certain ordinary differential equations and
obtain some exact solutions. However, many of their solutions are
equivalent to each other under the action of the known symmetry
transformations of the above equation.

The coupled two-dimensional cubic nonlinear  Schr\"{o}dinger
equations
$$ i\psi_t+c_1(\psi_{xx}+\psi_{yy})+(a_1|\psi|^2+b_1|\vf|^2)\psi=0,
\eqno(1.2)$$
$$ i\vf_t+c_2(\vf_{xx}+\vf_{yy})+(a_2|\psi|^2+b_2|\vf|^2)\vf=0
\eqno(1.3)$$ are used to describe interaction of electromagnetic
waves with different polarizations in nonlinear optics, where
$a_1,a_2,b_1,b_2,c_1$ and $c_2$ are real constants. Radhakrishnan
and Lakshmanan [RL1] used Painlev\'{e} analysis to find a Hirota
bilinearization of the above system of partial differential
equations and obtained bright and dark multiple soliton soutions.
They also generalized their results to the coupled nonlinear
Schr\"{o}dinger equations with higher-order effects in [RL2].
Gr\'{e}bert and Guillot [GG] construcetd periodic solutions of
coupled one-dimensional nonlinear Schr\"{o}dinger equations with
periodic boundary conditions in some resonance situations.
Moreover, Hioe and Salter [HS] found a connections between
Lam\'{e} functions and solutions of the above coupled equations.

In terms of real-valued functions, the above equations form systems
of nonlinear partial differential equations. Such systems can not be
solved exactly without pre-assumptions. We observe that the argument
functions of many known solutions for these equations are quadratic
in the spacial variables $x$ and $y$, in particular, those in [SEG].
Moreover, some of these solutions are actually equivalent to each
other under the Lie point symmetries of their corresponding
equations. These facts  motivate us to solve the above equations in
this paper by imposing the quadratic condition on the related
argument functions and using their symmetry transformations. More
complete families  of explicit exact solutions of this type with
multiple parameter functions are obtained. Many known interesting
solutions, such soliton ones, turn out to be special cases of our
solutions. Various  singular solutions and periodic solutions that
we obtain may reflect some important physical phenomena in practical
models. Our solutions can also be used to solve some boundary-value
problems. Below we give more details.

For convenience, we always assume that all the involved partial
derivatives of related functions always exist and we can change
orders of taking partial derivatives. We also use prime $'$ to
denote the derivative of any one-variable function.

It is known that the equation (1.1) is invariant under the
following known symmetric transformations:
$$T_1(\psi)=de^{d_3i}
\psi(d^2t+d_2,d(x\cos d_1+y\sin d_1),d(-x\sin d_1+y\cos
d_1)),\eqno(1.4)$$
$$T_2(\psi)=e^{[2(d_1x+d_3y)-(d_1^2+d_3^2)t]i/4c}
\psi(t,x-d_1t+d_2,y-d_3t+d_4),\eqno(1.5)$$ where
$d,d_1,d_2,d_3,d_4\in\mbb{R}$ with $d\neq 0$. In other words, the
above transformations transform one solution of (1.1) into another
solution. Our solutions contain all the solutions in [SEG] up to the
above transformations. In particular, our solutions with elliptic
functions were not given in [SEG]. Our approach is quite elementary
and accessible to large audiences  such as physicists and engineers.
For the reader's convenience, we list in this paper all the
solutions of the equation (1.1) found by our method although some of
them are known and obvious. This may help non-mathematicians to
apply the solutions of the Schr\"{o}dinger equation to their fields.
In fact, applying the transformations in (1.4) and (1.5) to any of
our solutions will yield more sophisticated one.

Similarly, we have the following known symmetric transformations
of the coupled equations (1.2) and (1.3):
$$T_1(\psi)=de^{d_3i}
\psi(d^2t+d_2,d(x\cos d_1+y\sin d_1),d(-x\sin d_1+y\cos
d_1)),\eqno(1.6)$$
$$T_1(\vf)=de^{d_4i}
\vf(d^2t+d_2,d(x\cos d_1+y\sin d_1),d(-x\sin d_1+y\cos
d_1));\eqno(1.7)$$
$$T_2(\psi)=e^{[2(d_1x+d_3y)-(d_1^2+d_3^2)t]i/4c_1}
\psi(t,x-d_1t+d_2,y-d_3t+d_4),\eqno(1.8)$$
$$T_2(\vf)=e^{[2(d_1x+d_3y)+(d_1^2+d_3^2)t]i/4c_2}
\vf(t,x-d_1t+d_2,y-d_3t+d_4);\eqno(1.9)$$
 where $d,d_1,d_2,d_3,d_4\in\mbb{R}$ with $d\neq 0$. In addition to the above
 symmetries, we
 also solve the coupled equations modulo the following symmetry:
 $$(\psi,a_1,b_1,c_1)\leftrightarrow
 (\vf,a_2,b_2,c_2).\eqno(1.10)$$
Again for the reader's convenience, we list in this paper all the
solutions of the coupled equations (1.2) and (1.3) found by our
method although some of them are known and obvious. For convenience,
we always assume that all the involved partial derivatives of
related functions always exist and we can change orders of taking
partial derivatives. We also use prime $'$ to denote the derivative
of any one-variable function.

In Section 2, we solve the Schr\"{o}dinger equation (1.1). In
Section 3, we use the results in Section 2 to solve the coupled
Schr\"{o}dinger equations (1.2) and (1.3).

\section{Exact Solutions of the  Schr\"{o}dinger
Equation}

In this section, we will present our quadratic-argument approach
to the two-dimensional cubic nonlinear  Schr\"{o}dinger equation
(1.1) and find more exact solutions than [SEG] in the modulo
sense.

 Write
$$\psi=\xi(t,x,y)e^{i\phi(t,x,y)},\eqno(2.1)$$
where $\xi$ and $\phi$ are real functions in $t,x,y$. Note
$$\psi_t=(\xi_t+i\xi\phi_t)e^{i\phi},\qquad
\psi_x=(\xi_x+i\xi\phi_x)e^{i\phi},\qquad
\psi_y=(\xi_y+i\xi\phi_y)e^{i\phi},\eqno(2.2)$$
$$\psi_{xx}=(\xi_{xx}-\xi\phi_x^2+i(2\xi_x\phi_x
+\xi\phi_{xx}))e^{i\phi},\;\;
\psi_{yy}=(\xi_{yy}-\xi\phi_y^2+i(2\xi_y\phi_y
+\xi\phi_{yy}))e^{i\phi}.\eqno(2.3)$$ So the equation (1.1) becomes
\begin{eqnarray*}\hspace{2cm}&
&i\xi_t-\phi_t\xi+a\xi^3
+c[\xi_{xx}+\xi_{yy}-\xi(\phi_x^2+\phi_y^2)\\
& &+i(2\xi_x\phi_x+2\xi_y
\phi_y+\xi(\phi_{xx}+\phi_{yy}))]=0,\hspace{5.4cm}
(2.4)\end{eqnarray*} equivalently,
$$\xi_t+c(2\xi_x \phi_x+2\xi_y
\phi_y+\xi(\phi_{xx}+\phi_{yy}))=0,\eqno(2.5)$$
$$-\xi[\phi_t+c(\phi_x^2+\phi_y^2)]
+c(\xi_{xx}+\xi_{yy})+a\xi^3=0.\eqno(2.6)$$

Note that it is very difficult to solve the above system without
pre-assumptions. We observe that the functions $\phi$ in all the
solutions of [SEG] are quadratic in $x$ and $y$. From the algebraic
characteristics of the above system of partial differential
equations, it is most affective to assume that $\phi$ is quadratic
in $x$ and $y$. After sorting case by case, we only have the
following four cases that lead us to exact solutions, modulo the
transformations in (1.4) and (1.5).\psp

{\it Case 1}. $\phi=\be(t)$ is a function of $t$.\psp

According to (2.5), $\xi_t=0$. Moreover, (2.6) becomes
$$-\be'\xi
+c(\xi_{xx}+\xi_{yy})+a\xi^3=0.\eqno(2.7)$$ So we take
$$\be=bt+d,\qquad b,d\in\mbb{R}.\eqno(2.8)$$
If $b=0$ and $ac<0$, modulo the transformations (1.4), we take $d=0$
and
 the following solutions:
 $$\xi=\frac{1}{x}\sqrt{-\frac{2c}{a}}\qquad\mbox{or}
 \qquad\sqrt{-\frac{c}{a(x^2+y^2)}}.
\eqno(2.9)$$ Next we assume $b\neq 0$. Modulo the transformation
(1.4), we can take $d=0$. Note that
$${(\tan s)'}'=2(\tan^3 s+\tan
s), \qquad{(\sec s)'}'=2\sec^3 s-\sec s,\eqno(2.10)$$
$${(\coth s)'}'=2(\coth^3s-\coth s),\qquad
{(\csch s)'}'=2\csch^3 s+\csch s.\eqno(2.11)$$ Denote Jacobi
elliptic functions
$$\sn s=\sn(s|m),\qquad\cn s=\cn(s|m),\qquad\dn s=\dn(s|m),\eqno(2.12)$$
where $m$ is the elliptic modulus (e.g., cf. [WG]). Then
$${(\sn s)'}'=2m^2\sn^3s-(1+m^2)\sn s,\eqno(2.13)$$
$${(\cn s)'}'=-2m^2\cn^2s+(2m^2-1)\cn s,\eqno(2.14)$$
$${(\dn s)'}'=-2\dn^3s+(2-m^2)\dn s.\eqno(2.15)$$
Moreover,
$$\lim_{m\rta 1}\sn s=\tanh s,\qquad \lim_{m\rta 1}\cn s=
\lim_{m\rta 1}\dn s=\sech s.\eqno(2.16)$$

Consider solutions modulo the transformation (1.4). If $ac<0$, we
have the following solutions:
$$\xi=\sqrt{-\frac{2c}{a}}\:\tan x,\qquad b=2c;\eqno(2.17)$$
$$\xi=\sqrt{-\frac{2c}{a}}\:\sec x,\qquad b=-c;\eqno(2.18)$$
$$\xi=\sqrt{-\frac{2c}{a}}\:\coth x,\qquad b=-2c;\eqno(2.19)$$
$$\xi=\sqrt{-\frac{2c}{a}}\:\csch x,\qquad b=c;\eqno(2.20)$$
$$\xi=m\sqrt{-\frac{2c}{a}}\:\sn x,\qquad b=-(1+m^2)c.\eqno(2.21)$$
When $ac>0$, we get the following solutions:
$$\xi=m\sqrt{\frac{2c}{a}}\:\cn x,\qquad b=(2m^2-1)c,\eqno(2.22)$$
$$\xi=\sqrt{\frac{2c}{a}}\:\dn x,\qquad b=(2-m^2)c.\eqno(2.23)$$
\pse

{\bf Theorem 2.1}.{\it Let $m\in\mbb{R}$. The following function
are solutions $\psi$ of the two-dimensional cubic nonlinear cubic
nonlinear Schr\"{o}dinger equation (1.1): if $ac<0$,
$$\frac{1}{x}\sqrt{-\frac{2c}{a}},
 \qquad\sqrt{-\frac{c}{a(x^2+y^2)}},\qquad
  e^{2cti}\sqrt{-\frac{2c}{a}}\:\tan x,\qquad
   e^{-cti}\sqrt{-\frac{2c}{a}}\sec x,
  \eqno(2.24)$$}
$$e^{-2cti}\sqrt{-\frac{2c}{a}}\:\coth x,\qquad
e^{cti}\sqrt{-\frac{2c}{a}}\:\csch x,\qquad
me^{-(1+m^2)cti}\sqrt{-\frac{2c}{a}}\sn x;\eqno(2.25)$$ {\it when}
$ac>0$,
$$me^{(2m^2-1)cti}\sqrt{\frac{2c}{a}}\:\cn x,\qquad e^{(2-m^2)cti}
\sqrt{\frac{2c}{a}}\:\dn x.\eqno(2.26)$$ \pse

{\bf Remark 2.2}. Although the above solution are simple, we can
obtain more sophisticated ones by applying the transformations (1.4)
and (1.5) to them. For instance, applying the transformation (1.4)
to the first solution in (2.24), we get a solution:
$$\psi=\frac{e^{d_2i}}{x\cos d_1+y\sin
d_1}\sqrt{-\frac{2c}{a}},\qquad d_1,d_2\in\mbb{R}.\eqno(2.27)$$
Applying the transformation (1.5) to the above solution, we obtain
another solution:
$$\psi=\frac{e^{[2(d_3x+d_4y)+(d_3^2+d_4^2)t+d_2]i/4c}}{(x-d_3t)\cos d_1
+(y-d_4t)\sin d_1+d_5}\sqrt{-\frac{2c}{a}},\qquad
d_1,d_2,d_3,d_4,d_5\in\mbb{R}.\eqno(2.28)$$ \pse

{\it Case 2}. $\phi=x^2/4ct+\be$ for some function $\be$ of
$t$.\psp

In this case, (2.5) becomes
$$\xi_t+\frac{x}{t}\xi_x +\frac{1}{2t}\xi=0.\eqno(2.29)$$
Thus
$$\xi=\frac{1}{\sqrt{t}}\zeta(u,y),\qquad u=\frac{x}{t},\eqno(2.30)$$
for some two-variable function $\zeta$. Now (2.6) becomes (2.7).
Note
$$
\xi_{xx}=t^{-5t/2}\zeta_{uu},\qquad\xi_{yy}=t^{-1/2}\zeta_{yy},\qquad\xi^3=t^{-3/2}\zeta^3.\eqno(2.31)$$
So (2.7) becomes
$$-\frac{\be'}{\sqrt{t}}\zeta+c(t^{-5t/2}\zeta_{uu}+t^{-1/2}\zeta_{yy})+at^{-3/2}\zeta^3
=0,\eqno(2.32)$$ whose coefficients of $t^{-3/2}$ force
 us to take
$$\xi=\frac{b}{\sqrt{t}},\qquad b\in\mbb{R}.\eqno(2.33)$$
Now (2.7) becomes
$$-\be'+\frac{ab^2}{t}=0\lra\be=ab^2\ln t\eqno(2.34)$$
modulo the transformation in (1.5).\psp

{\it Case 3}. $\phi=x^2/4ct+y^2/4c(t-d)+\be$ for some function
$\be$ of $t$ with $0\neq d\in\mbb{R}$. \psp

In this case, (2.5) becomes
$$\xi_t+\frac{x}{t}\xi_x +\frac{y}{t-d}\xi_y+
\left(\frac{1}{2t}+\frac{1}{2(t-d)}\right)\xi=0.\eqno(2.35)$$
 So we have:
$$\xi=\frac{1}{\sqrt{t(t-d)}}\zeta(u,v),\qquad
u=\frac{x}{t},\;v=\frac{y}{t-d}, \eqno(2.36)$$ for some two-variable
function $\zeta$.
 Again (2.6) becomes (2.7). Note
 $$\xi_{xx}=t^{-5/2}(t-c)^{-1/2}\zeta_{uu},\qquad\xi_{yy}=
 t^{-1/2}(t-c)^{-5/2}\zeta_{vv},\qquad\xi^3=t^{-3/2}(t-c)^{-3/2}\zeta^3.
 \eqno(2.37)$$
So (2.7) becomes
$$-\frac{\be'}{\sqrt{t(t-d)}}\zeta+c(t^{-5/2}(t-c)^{-1/2}\zeta_{uu}+t^{-1/2}(t-c)^{-5/2}\zeta_{vv})
+at^{-3/2}(t-c)^{-3/2}\zeta^3=0,\eqno(2.38)$$ whose coefficients of
$t^{-3/2}(t-c)^{-3/2}$ force
 us to take
$$\xi=\frac{b}{\sqrt{t(t-d)}},\qquad b\in\mbb{R}.\eqno(2.39)$$
Now (2.7) becomes
$$-\be'+\frac{ab^2}{t(t-d)}=0\lra\be=\frac{ab^2}{d}\ln
\frac{t-d}{t}\eqno(2.40)$$ modulo the transformation in (1.5).
 \psp

 {\bf Theorem 2.3}. {\it Let $b,d\in\mbb{R}$ with $d\neq 0$. The following function are solutions
$\psi$ of the two-dimensional cubic nonlinear cubic nonlinear
Schr\"{o}dinger equation:
$$bt^{ab^2i-1/2}e^{x^2i/4ct},\qquad
bt^{-ab^2i/d-1/2}(t-d)^{ab^2i/d-1/2}e^{x^2i/4ct+y^2i/4c(t-d)}
.\eqno(2.41)$$}\pse

{\bf Remark 2.4}. Applying (1.4) to the above first solution, we get
another solution
$$\psi=dd_1(d_1^2t+d_4)^{ad^2i-1/2}\exp\left(\frac{d_1^2
(x\cos d_2+y\sin d_2)^2}{4c(d_1^2t+d_4)}+d_3\right)i,\eqno(2.42)$$
for $d_1,d_2,d_3,d_4\in\mbb{R}$.  Moreover, we obtain a more
sophisticated solution:
\begin{eqnarray*}\hspace{2cm}\psi&=&dd_1(d_1^2t+d_4)^{ad^2i-1/2}\exp\frac{d_1^2
((x-d_5t)\cos d_2+(y-d_6t)\sin d_2+d_7)^2i}{4c(d_1^2t+d_4)}\\
& &\times\exp\left(\frac{2(d_5x+d_6y)+(d_5^2+d_6^2)t}{4c}
+d_3\right)i\hspace{4.2cm}(2.43)\end{eqnarray*} by applying the
transformation (1.5) to (2.42), where $b_r\in\mbb{R}$.\psp

{\it Case 4}. $\phi=(x^2+y^2)/4ct+\be$ for some function $\be$ of
$t$.\psp

Under our assumption, (2.5) becomes
$$\xi_t+\frac{x}{t}\xi_x +\frac{y}{t}\xi_y+
\frac{1}{t}\xi=0.\eqno(2.44)$$ Thus we have:
$$\xi=\frac{1}{t}\zeta(u,v),\qquad
u=\frac{x}{t},\;v=\frac{y}{t}, \eqno(2.45)$$ for some two-variable
function $\zeta$. Moreover, (2.6) becomes
$$-\be'\zeta+\frac{c}{t^2}(\zeta_{uu}+\zeta_{vv})
+\frac{a}{t^2}\zeta^3=0.\eqno(2.46)$$ An obvious solution is
$$\zeta=d,\qquad\be=-\frac{ad^2}{t},\qquad
d\in\mbb{R}.\eqno(2.47)$$ If $ac<0$, we have the simple following
solutions with $\be=0$:
$$\zeta=\frac{1}{\ell_1u+\ell_2v+\ell_3}\sqrt{-\frac{2c(\ell_1^2+\ell_2^2)}{a}}
\qquad\mbox{or}\qquad\sqrt{-\frac{c}{a((u-\ell_1)^2+(v-\ell_2)^2)}}\eqno(2.48)$$
for $\ell_1,\ell_2,\ell_3\in\mbb{R}$.

Next we assume
$$\be'=\frac{b}{t^2}\lra \be=-\frac{b}{t}\eqno(2.49)$$
modulo the transformation in (1.2), where $b$ is a real constant
to be determined. Suppose
$$\zeta=\Im(\varpi),\qquad \varpi=\ell_1u+\ell_2v+\ell_3\eqno(2.50)$$
for $\ell_1,\ell_2,\ell_3\in\mbb{R}$ such that $(\ell_1,\ell_2)\neq
(0,0)$. Then (2.46) is equivalent to:
$$-b\Im+c(\ell_1^2+\ell_2^2){\Im'}'+a\Im^3=0.\eqno(2.51)$$
According to (2.10), (2.11) and (2.13)-(2.15), we have the
following solutions: If $ac<0$, we have the following solutions:
$$\Im=\sqrt{-\frac{2c(\ell_1^2+\ell_2^2)}{a}}\:\tan \varpi,\qquad
b=2c(\ell_1^2+\ell_2^2);\eqno(2.52)$$
$$\Im=\sqrt{-\frac{2c(\ell_1^2+\ell_2^2)}{a}}\:\sec \varpi,\qquad
 b=-c(\ell_1^2+\ell_2^2);\eqno(2.53)$$
$$\Im=\sqrt{-\frac{2c(\ell_1^2+\ell_2^2)}{a}}\:\coth \varpi,\qquad
 b=-2c(\ell_1^2+\ell_2^2);\eqno(2.54)$$
$$\Im=\sqrt{-\frac{2c(\ell_1^2+\ell_2^2)}{a}}\:\csch\varpi,
\qquad b=c(\ell_1^2+\ell_2^2);\eqno(2.55)$$
$$\Im=m\sqrt{-\frac{2c(\ell_1^2+\ell_2^2)}{a}}\:\sn \varpi,
\qquad b=-(1+m^2)c(\ell_1^2+\ell_2^2).\eqno(2.56)$$ When $ac>0$, we
get the following solutions:
$$\Im=m\sqrt{\frac{2c(\ell_1^2+\ell_2^2)}{a}}\:\cn \varpi,\qquad
b=(2m^2-1)c(\ell_1^2+\ell_2^2),\eqno(2.57)$$
$$\Im=\sqrt{\frac{2c(\ell_1^2+\ell_2^2)}{a}}\:\dn \varpi,\qquad
 b=(2-m^2)c(\ell_1^2+\ell_2^2).\eqno(2.58)$$
\pse

{\bf Theorem 2.5}.{\it Let
$\ell_1,\ell_2,\ell_3,\ell_4,\ell_5,m\in\mbb{R}$ such that
$(\ell_1,\ell_2)\neq (0,0)$. The following functions are solutions
$\psi$ of the two-dimensional cubic nonlinear cubic nonlinear
Schr\"{o}dinger equation: if $ac<0$,
$$\frac{1}{\ell_1x+\ell_2y+\ell_3t}\sqrt{-\frac{2c(\ell_1^2+\ell_2^2)}{a}}\;e^{(x^2+y^2)i/4ct},
\eqno(2.59)$$
$$\sqrt{-\frac{c}{a((x-\ell_4t)^2+(y-\ell_5t)^2)}}\;e^{(x^2+y^2)i/4ct},\eqno(2.60)$$
$$\sqrt{-\frac{2c(\ell_1^2+\ell_2^2)}{a}}\:\frac{1}{t}\tan
\frac{\ell_1x+\ell_2y+\ell_3t}{t}\;\exp\left(\frac{x^2+y^2}{4ct}
-\frac{2c(\ell_1^2+\ell_2^2)}{t}\right)i,\eqno(2.61)$$
$$\sqrt{-\frac{2c(\ell_1^2+\ell_2^2)}{a}}\:\frac{1}{t}\sec
\frac{\ell_1x+\ell_2y+\ell_3t}{t}\;\exp\left(\frac{x^2+y^2}{4ct}
+\frac{c(\ell_1^2+\ell_2^2)}{t}\right)i,\eqno(2.62)$$
$$\sqrt{-\frac{2c(\ell_1^2+\ell_2^2)}{a}}\:\frac{1}{t}\coth
\frac{\ell_1x+\ell_2y+\ell_3t}{t}\;\exp\left(\frac{x^2+y^2}{4ct}
+\frac{2c(\ell_1^2+\ell_2^2)}{t}\right)i,\eqno(2.63)$$}
$$\sqrt{-\frac{2c(\ell_1^2+\ell_2^2)}{a}}\:\frac{1}{t}\csch
\frac{\ell_1x+\ell_2y+\ell_3t}{t}\;\exp\left(\frac{x^2+y^2}{4ct}
-\frac{c(\ell_1^2+\ell_2^2)}{t}\right)i,\eqno(2.64)$$
$$m\sqrt{-\frac{2c(\ell_1^2+\ell_2^2)}{a}}\:\frac{1}{t}\sn
\frac{\ell_1x+\ell_2y+\ell_3t}{t}\;\exp\left(\frac{x^2+y^2}{4ct}
+\frac{c(1+m^2)(\ell_1^2+\ell_2^2)}{t}\right)i;\eqno(2.65)$$ {\it
when} $ac>0$,
$$m\sqrt{\frac{2c(\ell_1^2+\ell_2^2)}{a}}\:\frac{1}{t}\cn
\frac{\ell_1x+\ell_2y+\ell_3t}{t}\;\exp\left(\frac{x^2+y^2}{4ct}
-\frac{c(2m^2-1)(\ell_1^2+\ell_2^2)}{t}\right)i,\eqno(2.66)$$
$$\sqrt{\frac{2c(\ell_1^2+\ell_2^2)}{a}}\:\frac{1}{t}\dn
\frac{\ell_1x+\ell_2y+\ell_3t}{t}\;\exp\left(\frac{x^2+y^2}{4ct}
-\frac{c(2-m^2)(\ell_1^2+\ell_2^2)}{t}\right)i.\eqno(2.67)$$ \pse

 {\bf Remark 2.6}. Applying the transformation (1.5) to the solution
(2.59), we get another solution:
$$\psi=\frac{e^{[2(d_1x+d_3y)+(d_1^2+d_3^2)t]i/4c}\sqrt{-\frac{2c(\ell_1^2+\ell_2^2)}{a}}}
{\ell_1(x-d_1t+d_2)+\ell_2(y-d_3t+d_4)+\ell_3t}
\;e^{((x-d_1t+d_2)^2+(y-d_3t+d_4)^2)i/4ct}, \eqno(2.68)$$ where
$d_1,d_2,d_3,d_4\in\mbb{R}$.

\section{Exact Solutions of the Coupled Equations}

In this section, we will use our results in previous section to find
exact solutions of  the coupled two-dimensional cubic nonlinear
Schr\"{o}dinger equations (1.2) and (1.3).

 Write
$$\psi=\xi(t,x,y)e^{i\phi(t,x,y)},\qquad \vf=\eta(t,x,y)e^{i\mu(t,x,y)}\eqno(3.1)$$
where $\xi,\phi,\eta$ and $\mu$ are real functions in $t,x,y$. As
the arguments in (2.1)-(2.6), the system (1.2) and (1.3) is
equivalent to the following system for real functions:
$$\xi_t+c_1(2\xi_x \phi_x+2\xi_y
\phi_y+\xi(\phi_{xx}+\phi_{yy}))=0,\eqno(3.2)$$
$$-\xi[\phi_t+c_1(\phi_x^2+\phi_y^2)]
+c_1(\xi_{xx}+\xi_{yy})+(a_1\xi^2+b_1\eta^2)\xi=0,\eqno(3.3)$$
$$\eta_t+c_2(2\eta_x \mu_x+2\eta_y
\mu_y+\eta(\mu_{xx}+\mu_{yy}))=0,\eqno(3.4)$$
$$-\eta[\mu_t+c_2(\mu_x^2+\mu_y^2)]
+c_2(\eta_{xx}+\eta_{yy})+(a_2\xi^2+b_2\eta^2)\eta=0.\eqno(3.5)$$
Based on our experience in last section, we will solve the above
system according to the following cases. For the convenience, we
always assume the conditions on the constants involved in an
expression such that it make sense. For instance, when we use
$\sqrt{d_1-d_2}$, we naturally assume $d_1\geq d_2$. \psp

{\it Case 1}. $(\phi,\mu)=(0,0)$ and $a_1b_2-a_2b_1\neq 0$.\psp

In this case, $\xi_t=\eta_t=0$ by (3.2) and (3.4). Moreover, (3.3)
and (3.5) become
$$c_1(\xi_{xx}+\xi_{yy})+(a_1\xi^2+b_1\eta^2)\xi=0,\qquad
c_2(\eta_{xx}+\eta_{yy})+(a_2\xi^2+b_2\eta^2)\eta=0.\eqno(3.6)$$
Assume
$$\xi=\frac{\iota_1}{x},\qquad\eta=\frac{\iota_2}{x}.\eqno(3.7)$$
Then (3.6) is equivalent to:
$$a_1\iota_1^2+b_1\iota_2^2+2c_1=0,\qquad
a_2\iota^2+b_2\iota^2+2c_2=0.\eqno(3.8)$$ Solving the above linear
algebraic equations for $\iota_1^2$ and $\iota_2^2$, we have:
$$\iota_1^2=\frac{2(b_1c_2-b_2c_1)}{a_1b_2-a_2b_1},\qquad\iota_2^2=
\frac{2(a_2c_1-a_1c_2)}{a_1b_2-a_2b_1}.\eqno(3.9)$$ Thus we have
the following solution
$$\xi=\frac{\es_1}{x}\sqrt{\frac{2(b_1c_2-b_2c_1)}{a_1b_2-a_2b_1}},\qquad
\eta=\frac{\es_2}{x}\sqrt{\frac{2(a_2c_1-a_1c_2)}{a_1b_2-a_2b_1}}
\eqno(3.10)$$ for $\es_1,\es_2\in\{1,-1\}$. Similarly, we have the
solution:
$$\xi=\es_1\sqrt{\frac{b_1c_2-b_2c_1}{(a_1b_2-a_2b_1)(x^2+y^2)}},\qquad
\eta=\es_2\sqrt{\frac{a_2c_1-a_1c_2}{(a_1b_2-a_2b_1)(x^2+y^2)}}.\eqno(3.11)$$
\psp

{\it Case 2}. $(\phi,\mu)=(k_1t,k_2t)$ with $k_1,k_2\in\mbb{R}$.
\psp

Again we have $\xi_t=\eta_t=0$ by (3.2) and (3.4). Moreover, (3.3)
and (3.5) become
$$-k_1\xi+c_1(\xi_{xx}+\xi_{yy})+(a_1\xi^2+b_1\eta^2)\xi=0,\;\;
-k_2\eta+c_2(\eta_{xx}+\eta_{yy})+(a_2\xi^2+b_2\eta^2)\eta=0.\eqno(3.12)$$

First we assume $a_1b_2-a_2b_1\neq 0$ and
$$\xi=\iota_1\Im(x),\qquad \eta=\iota_2\Im(x).\eqno(3.13)$$
Then (3.12) becomes
$$-k_1\Im+c_1{\Im'}'+(a_1\iota_1^2+b_1\iota_2^2)\Im^3=0,\qquad
-k_2\Im+c_2{\Im'}'+(a_2\iota^2+b_2\iota^2)\Im^3=0.\eqno(3.14)$$
 According to (2.10) and (2.11), when $\Im=\tan
x,\;\sec x,\;\coth x$ and $\csch x$, we always have
$$a_1\iota_1^2+b_1\iota_2^2+2c_1=0,\qquad
a_2\iota^2+b_2\iota^2+2c_2=0.\eqno(3.15)$$  Thus for
$\es_1,\es_2\in\{1,-1\}$, we have the following solutions:
$$\xi=\es_1\sqrt{\frac{2(b_1c_2-b_2c_1)}{a_1b_2-a_2b_1}}\:\tan x,\;\;
\eta=\es_2\sqrt{\frac{2(a_2c_1-a_1c_2)}{a_1b_2-a_2b_1}}\:\tan
x,\;\; (k_1,k_2)=2(c_1,c_2);\eqno(3.16)$$
$$\xi=\es_1\sqrt{\frac{2(b_1c_2-b_2c_1)}{a_1b_2-a_2b_1}}\:\sec x,\;\;
\eta=\es_2\sqrt{\frac{2(a_2c_1-a_1c_2)}{a_1b_2-a_2b_1}}\:\sec
x,\;\; (k_1,k_2)=-(c_1,c_2);\eqno(3.17)$$
$$\xi=\es_1\sqrt{\frac{2(b_1c_2-b_2c_1)}{a_1b_2-a_2b_1}}\:\coth
x,\;\;
\eta=\es_2\sqrt{\frac{2(a_2c_1-a_1c_2)}{a_1b_2-a_2b_1}}\:\coth
x\eqno(3.18)$$ and $(k_1,k_2)=-2(c_1,c_2);$
$$\xi=\es_1\sqrt{\frac{2(b_1c_2-b_2c_1)}{a_1b_2-a_2b_1}}\:\csch x,\;\;
\eta=\es_2\sqrt{\frac{2(a_2c_1-a_1c_2)}{a_1b_2-a_2b_1}}\:\csch
x,\;\;(k_1,k_2)=(c_1,c_2).\eqno(3.19)$$ Similarly, (2.13)-(2.15)
give us the following solutions:
$$\xi=m\es_1\sqrt{\frac{2(b_1c_2-b_2c_1)}{a_1b_2-a_2b_1}}\;\sn x,\qquad
\eta=m\es_2\sqrt{\frac{2(a_2c_1-a_1c_2)}{a_1b_2-a_2b_1}}\;\sn x
\eqno(3.20)$$ and $(k_1,k_2)=-(1+m^2)(c_1,c_2);$
$$\xi=m\es_1\sqrt{\frac{2(b_2c_1-b_1c_2)}{a_1b_2-a_2b_1}}\;\cn x\qquad
\eta=m\es_2\sqrt{\frac{2(a_1c_2-a_2c_1)}{a_1b_2-a_2b_1}}\;\cn x
\eqno(3.21)$$ and $(k_1,k_2)=(2m^2-1)(c_1,c_2);$
$$\xi=\es_1\sqrt{\frac{2(b_2c_1-b_1c_2)}{a_1b_2-a_2b_1}}\;\dn
x,\qquad
\eta=\es_2\sqrt{\frac{2(a_1c_2-a_2c_1)}{a_1b_2-a_2b_1}}\;\dn
x\eqno(3.22)$$ and $(k_1,k_2)=(2-m^2)(c_1,c_2).$

If $(a_1,b_1)= a_1(1,d^2)$ and $(a_2,b_2)=a_2(1,d^2)$ with
$d\in\mbb{R}$, we have the following solution of (3.12):
$$\xi=d\ell\sin x,\qquad \eta=\ell\cos x,\qquad
(k_1,k_2)=(a_1(d\ell)^2-c_1,a_2(d\ell)^2-c_2)\eqno(3.23)$$ for
$\ell\in\mbb{R}$.
 When
$(a_1,b_1)= a_1(1,-d^2)$ and $(a_2,b_2)=a_2(1,-d^2)$ with
$d\in\mbb{R}$, we get the solution:
$$\xi=d\ell\cosh x,\qquad \eta=\ell\sinh x,\qquad
(k_1,k_2)=(a_1(d\ell)^2+c_1,a_2(d\ell)^2+c_2).\eqno(3.24)$$ In
summary, we have the following theorem. \psp

{\bf Theorem 3.1}. {\it Let $d,\ell,m\in\mbb{R}$ and let
$\es_1,\es_2\in\{1,-1\}$. If $a_1b_2-a_2b_1\neq 0$, we have the
following solutions of the coupled two-dimensional cubic nonlinear
Schr\"{o}dinger equations (1.2) and (1.3):
$$\psi=\frac{\es_1}{x}\sqrt{\frac{2(b_1c_2-b_2c_1)}{a_1b_2-a_2b_1}},\qquad
\vf=\frac{\es_2}{x}\sqrt{\frac{2(a_2c_1-a_1c_2)}{a_1b_2-a_2b_1}};
\eqno(3.25)$$
$$\psi=\es_1\sqrt{\frac{b_1c_2-b_2c_1}{(a_1b_2-a_2b_1)(x^2+y^2)}},\qquad
\vf=\es_2\sqrt{\frac{a_2c_1-a_1c_2}{(a_1b_2-a_2b_1)(x^2+y^2)}};\eqno(3.26)$$
$$\psi=\es_1\sqrt{\frac{2(b_1c_2-b_2c_1)}{a_1b_2-a_2b_1}}\:e^{2c_1ti}\tan x,\;\;
\vf=\es_2\sqrt{\frac{2(a_2c_1-a_1c_2)}{a_1b_2-a_2b_1}}\:e^{2c_2ti}\tan
x;\eqno(3.27)$$
$$\psi=\es_1\sqrt{\frac{2(b_1c_2-b_2c_1)}{a_1b_2-a_2b_1}}\:e^{-c_1ti}\sec x,\;\;
\vf=\es_2\sqrt{\frac{2(a_2c_1-a_1c_2)}{a_1b_2-a_2b_1}}\:e^{-c_2ti}\sec
x;\eqno(3.28)$$
$$\psi=\es_1\sqrt{\frac{2(b_1c_2-b_2c_1)}{a_1b_2-a_2b_1}}\:e^{-2c_1ti}\coth x,\;\;
\vf=\es_2\sqrt{\frac{2(a_2c_1-a_1c_2)}{a_1b_2-a_2b_1}}\:e^{-2c_2ti}\coth
x;\eqno(3.29)$$}
$$\psi=\es_1\sqrt{\frac{2(b_1c_2-b_2c_1)}{a_1b_2-a_2b_1}}\:e^{c_1ti}\csch
x,\qquad
\vf=\es_2\sqrt{\frac{2(a_2c_1-a_1c_2)}{a_1b_2-a_2b_1}}\:e^{c_2ti}\csch
x;\eqno(3.30)$$
$$\psi=m\es_1\sqrt{\frac{2(b_1c_2-b_2c_1)}{a_1b_2-a_2b_1}}\;e^{-(1+m^2)c_1ti}
\sn x,\eqno(3.31)$$
$$\vf=m\es_2\sqrt{\frac{2(a_2c_1-a_1c_2)}{a_1b_2-a_2b_1}}\;e^{-(1+m^2)c_2ti}\sn
x; \eqno(3.32)$$
$$\psi=m\es_1\sqrt{\frac{2(b_2c_1-b_1c_2)}{a_1b_2-a_2b_1}}\;e^{(2m^2-1)c_1ti}
\cn x,\eqno(3.33)$$
$$\vf=m\es_2\sqrt{\frac{2(a_1c_2-a_2c_1)}{a_1b_2-a_2b_1}}\;e^{(2m^2-1)c_2ti}\cn
x;\eqno(3.34)$$
$$\psi=\es_1\sqrt{\frac{2(b_2c_1-b_1c_2)}{a_1b_2-a_2b_1}}\;e^{(2-m^2)c_1ti}
\dn x,\;\;
\vf=\es_2\sqrt{\frac{2(a_1c_2-a_2c_1)}{a_1b_2-a_2b_1}}\;e^{(2-m^2)c_1ti}\dn
x.\eqno(3.35)$$ {\it If $(a_1,b_1)= a_1(1,d^2)$ and
$(a_2,b_2)=a_2(1,d^2)$},
$$\psi=d\ell e^{(a_1(d\ell)^2-c_1)ti}\sin x,\qquad\vf=\ell e^{(a_2(d\ell)^2
-c_2)ti} \cos x.\eqno(3.36)$$ {\it When $(a_1,b_1)= a_1(1,-d^2)$
and $(a_2,b_2)=a_2(1,-d^2)$},
$$\psi=d\ell e^{(a_1(d\ell)^2+c_1)ti}\cosh x,\qquad
\eta=\ell e^{(a_2(d\ell)^2+c_2)ti}\sinh x.\eqno(3.37)$$ \pse

{\bf Remark 3.2}. Applying the symmetric transformations
(1.6)-(1.9) to the above solutions, we can get more sophisticated
ones. For instance, applying $T_1$ in (1.6) and (1.7) to (3.25)
and (3.37), we get
$$\psi=\frac{\es_1e^{d_2i}}{x\cos d_1+y\sin d_1}
\sqrt{\frac{2(b_1c_2-b_2c_1)}{a_1b_2-a_2b_1}},\;\;
\vf=\frac{\es_2e^{d_3i}}{x\cos d_1+y\sin
d_1}\sqrt{\frac{2(a_2c_1-a_1c_2)}{a_1b_2-a_2b_1}}; \eqno(3.38)$$
and
$$\psi=
dd_2\ell e^{[(a_1(d\ell)^2+c_1)d_2^2t+d_3]i}\cosh d_2(x\cos
d_1+y\sin d_1),\eqno(3.39)$$ $$ \eta=d_2\ell
e^{[(a_2(d\ell)^2+c_2)d_2^2t+d_4]i}\sinh d_2(x\cos d_1+y\sin
d_1).\eqno(3.40)$$ Applying $T_2$ in (1.8) and (1.9) to (3.26), we
obtain:
$$\psi=\es_1e^{[2(d_1x+d_3y)+(d_1^2+d_3^2)t]i/4c_1}
\sqrt{\frac{b_1c_2-b_2c_1}{(a_1b_2-a_2b_1)
((x-d_1t+d_2)^2+(y-d_2t+d_4)^2)}},\eqno(3.41)$$
$$\vf=\es_2e^{[2(d_1x+d_3y)+(d_1^2+d_3^2)t]i/4c_2}
\sqrt{\frac{a_2c_1-a_1c_2}{(a_1b_2-a_2b_1)((x-d_1t+d_2)^2+(y-d_2t+d_4)^2)}
}.\eqno(3.42)$$ \pse

{\it Case 3}. $\phi=x^2/4c_1t+\be_1$ and
$\mu=(x-d)^2/4c_2(t-\ell)+\be_2$ or $\mu=y^2/4c_2(t-\ell)+\be_2$
 for some functions $\be_1$ and $\be_2$ of $t$ and real constants
 $d$ and $\ell$.\psp

First we assume $\mu=(x-d)^2/4c_2(t-\ell)+\be_2$. Then (3.2) and
(3.4) become
$$\xi_t+\frac{x}{t}\xi_x+\frac{1}{2t}\xi=0,\qquad
 \eta_t+\frac{x-d}{t-\ell}\eta_x+\frac{1}{2(t-\ell)}\eta=0.
\eqno(3.43)$$ Thus
$$\xi=\frac{1}{\sqrt{t}}\hat\xi(t^{-1}x,y),\qquad\eta=
\frac{1}{\sqrt{t-\ell}}\hat\eta((t-\ell)^{-1}(x-d),y)\eqno(3.44)$$
for some two-variable functions $\hat\xi$ and $\hat\eta$. On the
other hand, (3.3) and (3.5) become:
$$-\be_1'\xi+c_1(\xi_{xx}+\xi_{yy})+(a_1\xi^2+b_1\eta^2)\xi=0,\eqno(3.45)$$
$$-\be_2'\eta+c_2(\eta_{xx}+\eta_{yy})+(a_2\xi^2+b_2\eta^2)\eta=0.
\eqno(3.46)$$

As (2.31)-(2.33), the above two equations force us to take
$$\xi=\frac{k_1}{\sqrt{t}},\qquad\eta=
\frac{k_2}{\sqrt{t-\ell}}.\eqno(3.47)$$ So (3.45) and (3.46) are
implied by the equations:
$$\be_1'=\frac{a_1k_1^2}{t}+\frac{b_1k_2^2}{t-\ell},\qquad
\be_2'=\frac{a_2k_1^2}{t}+\frac{b_2k_2^2}{t-\ell}.\eqno(3.48)$$
Modulo the transformation (1.6) and (1.7), we take
$$\be_1=a_1k_1^2\ln t+b_1k_2^2\ln(t-\ell),\qquad
\be_2=a_2k_1^2\ln t+b_2k_2^2\ln(t-\ell).\eqno(3.49)$$ Exact same
approach holds for $\mu=y^2/4c_2(t-\ell)+\be_2$.\psp

{\bf Theorem 3.3}. {\it Let $d,\ell,k_1,k_2\in\mbb{R}$. We have
the following solutions of the coupled two-dimensional cubic
nonlinear Schr\"{o}dinger equations (1.2) and (1.3)}:
$$\psi=k_1t^{a_1k_1^2i-1/2}(t-\ell)^{b_1k_2^2i}e^{x^2i/2c_1t},\qquad
\vf=k_2t^{a_2k_1^2i}(t-\ell)^{b_2k_2^2i-1/2}e^{(x-d)^2i/2c_2(t-\ell)};
\eqno(3.50)$$
$$\psi=k_1t^{a_1k_1^2i-1/2}(t-\ell)^{b_1k_2^2i}e^{x^2i/2c_1t},\qquad
\vf=k_2t^{a_2k_1^2i}(t-\ell)^{b_2k_2^2i-1/2}e^{y^2i/2c_2(t-\ell)}.
\eqno(3.51)$$\pse

{\it Case 4}. $\phi=x^2/4c_1t+\be_1$ and
$\mu=(x-d)^2/4c_2(t-d_1)+y^2/4c_2(t-d_2)+\be_2$
 for some functions $\be_1$ and $\be_2$ of $t$ and real constants
 $d,d_1$ and $d_2$.
 \psp

In this case,  (3.2) and (3.4) become
$$\xi_t+\frac{x}{t}\xi_x+\frac{1}{2t}\xi=0,\;\;\eta_t+\frac{x-d}{t-d_1}\eta_x +\frac{y}{t-d_2}\eta_y+
\left(\frac{1}{2(t-d_1)}+\frac{1}{2(t-d_2)}\right)\xi=0.
\eqno(3.52)$$ Thus
$$\xi=\frac{1}{\sqrt{t}}\hat\xi(t^{-1}x,y),\qquad\eta=
\frac{1}{\sqrt{(t-d_1)(t-d_2)}}\hat\eta((t-d_1)^{-1}(x-d),
(t-d_2)^{-1}y)\eqno(3.53)$$ for some two-variable functions
$\hat\xi$ and $\hat\eta$. Again (3.3) and (3.5) become (3.45) and
(3.46), respectively. Moreover, they force us to take
$$\xi=\frac{k_1}{\sqrt{t}},\qquad\eta=
\frac{k_2}{\sqrt{(t-d_1)(t-d_2)}}.\eqno(3.54)$$ So (3.3) and (3.5)
are implied by the equations:
$$\be_1'=\frac{a_1k_1^2}{t}+\frac{b_1k_2^2}{(t-d_1)(t-d_2)},\qquad
\be_2'=\frac{a_2k_1^2}{t}+\frac{b_2k_2^2}{(t-d_1)(t-d_2)}.
\eqno(3.55)$$ Modulo the transformation (1.6) and (1.7), we get
$$\be_1=a_1k_1^2\ln t+\frac{b_1k_2^2}{d_2-d_1}\ln\frac
{t-d_1}{t-d_2},\qquad\be_2=a_2k_1^2\ln
t+\frac{b_2k_2^2}{d_2-d_1}\ln\frac {t-d_1}{t-d_2}\eqno(3.56)$$ if
$d_1\neq d_2$, and
$$\be_1=a_1k_1^2\ln t-\frac{b_1k_2^2}{t-d_1},
,\qquad\be_2=a_2k_1^2\ln t-\frac{b_2k_2^2}{t-d_1}\eqno(3.57)$$
when $d_1=d_2$.\psp

{\bf Theorem 3.4}. {\it Let $d_1,d_2,k_1,k_2\in\mbb{R}$ such that
$d_1\neq d_2$. We have the following solutions of the coupled
two-dimensional cubic nonlinear Schr\"{o}dinger equations (1.2)
and (1.3)}:
$$\psi=k_1t^{a_1k_1^2i-1/2}(t-d_1)^{b_1k_2^2(\ell_2-d_1)^{-1}i}
(t-d_2)^{-b_1k_2^2(\ell_2-d_1)^{-1}i}e^{x^2i/4c_1t},\eqno(3.58)$$
\begin{eqnarray*}\hspace{2cm}\vf&=&k_2t^{a_2k_1^2i}(t-d_1)^{b_1k_2^2(\ell_2-d_1)^{-1}i-1/2}
(t-d_2)^{-b_1k_2^2(\ell_2-d_1)^{-1}i-1/2}\\
& &\times\exp\left(\frac{(x-d)^2i} {4c_2(t-d_1)}+\frac{y^2i}
{4c_2(t-d_1)}\right);\hspace{5.2cm}(3.59)\end{eqnarray*}
$$\psi=k_1t^{a_1k_1^2i-1/2}\exp\left(\frac{x^2i} {4c_1t}
-\frac{b_1k_2^2i}{t-d_1}\right),\eqno(3.60)$$
$$\vf=\frac{k_2t^{a_2k_1^2i}}{t-d_1}
\exp\frac{((x-d)^2+y^2-4c_2b_2k_2^2)i} {4c_2(t-d_1)}.\eqno(3.61)$$
\pse

{\it Case 5}. For $\ell_1,\ell_2,\ell,d_1,d_2\in\mbb{R}$ and
functions $\be_1,\be_2$ of $t$,
$$\phi=\frac{x^2}{4c_1t}+\frac{y^2}{4c_1(t-\ell)}+\be_1,\qquad
\mu=\frac{(x-d_1)^2}{4c_2(t-\ell_1)}+\frac{(y-d_2)^2}{4c_1(t-\ell_2)}+\be_2.
\eqno(3.62)$$\pse

As in the above case, we get
$$\xi=\frac{k_1}{\sqrt{t(t-\ell)}},\qquad\eta=
\frac{k_2}{\sqrt{(t-\ell_1)(t-\ell_2)}}.\eqno(3.63)$$ So (3.3) and
(3.5) are implied by the equations:
$$\be_1'=\frac{a_1k_1^2}{t(t-\ell)}+\frac{b_1k_2^2}{(t-\ell_1)(t-\ell_2)},\qquad
\be_2'=\frac{a_2k_1^2}{t}+\frac{b_2k_2^2}{(t-\ell_1)(t-\ell_2)}.
\eqno(3.64)$$ Modulo the transformation (1.6) and (1.7), we have
$$\be_1=\frac{a_1k_1^2}{\ell}\ln \frac{t-\ell}{t}+\frac{b_1k_2^2}{\ell_2-\ell_1}\ln\frac
{t-\ell_1}{t-\ell_2},\qquad\be_2=\frac{a_2k_1^2}{\ell}\ln
\frac{t-\ell}{t}+\frac{b_2k_2^2}{\ell_2-\ell_1}\ln\frac
{t-\ell_1}{t-\ell_2}\eqno(3.65)$$ if $\ell\neq 0$ and
$\ell_1\neq\ell_2$;
$$\be_1=-\frac{a_1k_1^2}{t}+\frac{b_1k_2^2}{\ell_2-\ell_1}\ln\frac
{t-\ell_1}{t-\ell_2},\qquad\be_2=-\frac{a_2k_1^2}{t}+\frac{b_2k_2^2}{\ell_2-\ell_1}\ln\frac
{t-\ell_1}{t-\ell_2}\eqno(3.66)$$ when $\ell=0$  and
$\ell_1\neq\ell_2$;
$$\be_1=\frac{a_1k_1^2}{t}-\frac{b_1k_2^2}{t-\ell_1},
,\qquad\be_2=\frac{a_2k_1^2}{t}
t-\frac{b_2k_2^2}{t-\ell_1}\eqno(3.67)$$ if $\ell=0$ and
$\ell_1=\ell_2$. Therefore, we obtain:\psp

{\bf Theorem 3.5}. {\it Let
$\ell_1,\ell_2,\ell,d_1,d_2,k_1,k_2\in\mbb{R}$ such that $\ell\neq
0$ and $\ell_1\neq\ell_2$. We have the following solutions of the
coupled two-dimensional cubic nonlinear Schr\"{o}dinger equations
(1.2) and (1.3)}:
$$\psi=\frac{k_1}{t}
\exp\left(\frac{(x^2+y^2-4c_1a_1k_1^2)i}
{4c_1t}-\frac{b_1k_2^2i}{t-\ell_1}\right),\eqno(3.68)$$
$$\vf=\frac{k_2}{t-\ell_1}
\exp\left(\frac{((x-d_1)^2+(y-d_2)^2-4c_2b_2k_2^2)i}
{4c_2(t-\ell_1)}-\frac{a_2k_1^2i}{t}\right);\eqno(3.69)$$
$$\psi=\frac{k_1(t-\ell_1)^{b_1k_2^2(\ell_2-\ell_1)^{-1}i}
(t-\ell_2)^{-b_1k_2^2(\ell_2-\ell_1)^{-1}i}}{t}
\exp\frac{(x^2+y^2-4c_1a_1k_1^2)i} {4c_1t},\eqno(3.70)$$
\begin{eqnarray*}\hspace{2cm}\vf&=&k_2(t-\ell_1)^{b_2k_2^2(\ell_2-\ell_1)^{-1}i-1/2}
(t-\ell_2)^{-b_2k_2^2(\ell_2-\ell_1)^{-1}i-1/2}
\\ & &\times\exp\left(\frac{(x-d_1)^2i}{4c_2(t-\ell_1)}+\frac{(y-d_2)^2i}{4c_2(t-\ell_2)}
-\frac{a_2k_1^2i}{t}\right);\hspace{3.7cm}(3.71)\end{eqnarray*}
\begin{eqnarray*}\hspace{1.9cm}\psi&=&k_1t^{-a_1k_1^2\ell^{-1}i-1/2}
(t-\ell)^{a_1k_1^2\ell^{-1}i-1/2}
(t-\ell_1)^{b_1k_2^2(\ell_2-\ell_1)^{-1}i}\\ & &\times
(t-\ell_2)^{-b_1k_2^2(\ell_2-\ell_1)^{-1}i}
\exp\left(\frac{x^2i}{4c_1t}+\frac{y^2i}{4c_1(t-\ell)}
\right),\hspace{3cm}(3.72)\end{eqnarray*}
\begin{eqnarray*}\hspace{1cm}\vf&=&k_2t^{-a_2k_1^2\ell^{-1}i}
(t-\ell)^{a_2k_1^2\ell^{-1}i}
(t-\ell_1)^{b_2k_2^2(\ell_2-\ell_1)^{-1}i-1/2}\\ & &\times
(t-\ell_2)^{-b_2k_2^2(\ell_2-\ell_1)^{-1}i-1/2}
\exp\left(\frac{(x-d_1)^2i}{4c_2(t-\ell_1)}+
\frac{(y-d_2)^2i}{4c_2(t-\ell_2)}
\right).\hspace{2cm}(3.73)\end{eqnarray*} \pse

{\it Case 6}. For  two functions $\be_1,\be_2$ of $t$,
$$\phi=\frac{x^2+y^2}{4c_1t}+\be_1,\qquad
\mu=\frac{x^2+y^2}{4c_2t}+\be_2. \eqno(3.74)$$\pse

As in Case 4, (3.2) and (3.4) imply
$$\xi=\frac{1}{t}\hat\xi(u,v),\qquad
\eta=\frac{1}{t}\hat\eta(u,v),\qquad
u=\frac{x}{t},\;v=\frac{y}{t}.\eqno(3.75)$$ Moreover, (3.3) and
(3.5) become
$$-\be_1'\hat\xi
+\frac{c_1}{t^2}(\hat\xi_{uu}+\hat\xi_{vv})+
\frac{1}{t^2}(a_1\hat\xi^2+b_1\hat\eta^2)\hat\xi=0,\eqno(3.76)$$
$$-\be_2'\hat\eta
+\frac{c_2}{t^2}(\hat\eta_{uu}+\hat\eta_{vv})+
\frac{1}{t^2}(a_2\hat\xi^2+b_2\hat\eta^2)\hat\eta=0.\eqno(3.77)$$
To solve the above system, we assume
$$\be_1=-\frac{k_1}{t},\qquad \be_2=-\frac{k_2}{t},\qquad
k_1,k_2\in\mbb{R}.\eqno(3.78)$$ Then (3.77) and (3.78) are
equivalent to:
$$-k_1\hat\xi
+c_1(\hat\xi_{uu}+\hat\xi_{vv})+
(a_1\hat\xi^2+b_1\hat\eta^2)\hat\xi=0,\eqno(3.79)$$
$$-k_2\hat\eta
+c_2(\hat\eta_{uu}+\hat\eta_{vv})+(a_2\hat\xi^2+b_2\hat\eta^2)
\hat\eta=0.\eqno(3.80)$$

For $\ell_1,\ell_2,\ell_3\in\mbb{R}$, we set
$$\varpi=\ell_1u+\ell_2v+\ell_3.\eqno(3.81)$$
If $(a_1,b_1)= a_1(1,d^2)$ and $(a_2,b_2)=a_2(1,d^2)$ with
$d\in\mbb{R}$, we have the following solution:
$$\hat\xi=d\ell\sin\varpi,\;\; \hat\eta=\ell\cos\varpi,\;\;
(k_1,k_2)=(a_1(d\ell)^2-c_1(\ell_1^2+\ell_2^2),a_2(d\ell)^2-
c_2(\ell_1^2+\ell_2^2))\eqno(3.82)$$ for $\ell\in\mbb{R}$.
 When
$(a_1,b_1)= a_1(1,-d^2)$ and $(a_2,b_2)=a_2(1,-d^2)$ with
$d\in\mbb{R}$, we get the solution:
$$\hat\xi=d\ell\cosh\varpi,\;\; \hat\eta=\ell\sinh\varpi,\;\;
(k_1,k_2)=(c_1(\ell_1^2+\ell_2^2)+a_1(d\ell)^2,
c_2(\ell_1^2+\ell_2^2)+a_2(d\ell)^2)\eqno(3.83)$$ for
$\ell\in\mbb{R}$.\psp

{\bf Theorem 3.6}. {\it For $d,\ell,\ell_1,\ell_2,\ell_3\in\mbb{R}$,
we have the following solutions of the coupled two-dimensional cubic
nonlinear Schr\"{o}dinger equations (1.2) and (1.3)}:
$$\psi=\frac{d\ell\sin\frac{\ell_1x+\ell_2y+\ell_3t}{t}}
{t}\exp\left(\frac{x^2+y^2}{4c_1t}+\frac{c_1(\ell_1^2+\ell_2^2)
-a_1(d\ell)^2}{t}\right)i,\eqno(3.84)$$
$$\vf=\frac{\ell\cos\frac{\ell_1x+\ell_2y+\ell_3t}{t}}
{t}\exp\left(\frac{x^2+y^2}{4c_2t}+\frac{c_2(\ell_1^2+\ell_2^2)
-a_2(d\ell)^2}{t}\right)i\eqno(3.85)$$ {\it if $(a_1,b_1)=
a_1(1,d^2)$ and $(a_2,b_2)=a_2(1,d^2)$};
$$\psi=\frac{d\ell\cosh\frac{\ell_1x+\ell_2y+\ell_3t}{t}}
{t}\exp\left(\frac{x^2+y^2}{4c_1t}-\frac{c_1(\ell_1^2+\ell_2^2)
+a_1(d\ell)^2}{t}\right)i,\eqno(3.86)$$
$$\vf=\frac{\ell\sinh\frac{\ell_1x+\ell_2y+\ell_3t}{t}}
{t}\exp\left(\frac{x^2+y^2}{4c_2t}-\frac{c_2(\ell_1^2+\ell_2^2)
+a_2(d\ell)^2}{t}\right)i\eqno(3.87)$$ {\it when $(a_1,b_1)=
a_1(1,-d^2)$ and $(a_2,b_2)=a_2(1,-d^2)$}.\psp

Finally, we assume $a_1b_2-a_2b_1\neq 0$. Recall the notion in
(3.81). Taking $k_1=k_2=0$, the we have the following solutions of
the system of (3.79) and (3.80):
$$\hat\xi=\frac{1}{\varpi}\sqrt{\frac{2(b_1c_2-b_2c_1)(\ell_1^2+\ell_2^2)}{a_1b_2-a_2b_1}},\qquad
\hat\eta=\frac{1}{\varpi}\sqrt{\frac{2(a_2c_1-a_1c_2)(\ell_1^2+\ell_2^2)}
{a_1b_2-a_2b_1}}\eqno(3.88)$$ and
$$\hat\xi=\sqrt{\frac{b_1c_2-b_2c_1}{(a_1b_2-a_2b_1)((u-d_1)^2+(v-d_2)^2)}},
\eqno(3.89)$$
$$\hat\eta=\sqrt{\frac{a_2c_1-a_1c_2}{(a_1b_2-a_2b_1)((u-d_1)^2+(v-d_2)^2)}}
\eqno(3.90)$$ for $\ell_1,\ell_2\in\mbb{R}$. In general, we assume
$$\hat\xi=\iota_1\Im(\varpi),\qquad
\hat\eta=\iota_2\Im(\varpi)\eqno(3.91)$$ for some one-variable
function $\Im$ and $k_1,k_2\in\mbb{R}$. Then (3.79) and (3.80)
implied by
$$-k_1\Im +c_1(\ell_1^2+\ell_2^2){\Im'}'+
(a_1\iota_1^2+b_1\iota_2^2)\Im^3=0,\eqno(3.92)$$
$$-k_2\Im +c_2(\ell_1^2+\ell_2^2){\Im'}'+
(a_2\iota_1^2+b_2\iota_2^2)\Im^3=0.\eqno(3.93)$$

Recall $\es_1,\es_2\in\{1,-1\}$. Again by (2.10)-(2.15), Case 1
and Case 2, we have:
$$\iota_1=\es_1\sqrt{\frac{2(b_1c_2-b_2c_1)
(\ell_1^2+\ell_2^2)}{a_1b_2-a_2b_1}},\;\;
\iota_2=\es_2\sqrt{\frac{2(a_2c_1-a_1c_2)(\ell_1^2+\ell_2^2)}{a_1b_2-a_2b_1}}
\eqno(3.94)$$ with $(\Im,k_1,k_2)$ as follows:
$$(\tan\varpi,2c_1(\ell_1^2+\ell_2^2),2c_2(\ell_1^2+\ell_2^2)),\;\;
(\sec\varpi,-c_1(\ell_1^2+\ell_2^2),-c_2(\ell_1^2+\ell_2^2)),\eqno(3.95)$$
$$(\coth\varpi,-2c_1(\ell_1^2+\ell_2^2),-2c_2(\ell_1^2+\ell_2^2)),\;\;
(\csch\varpi,c_1(\ell_1^2+\ell_2^2),c_2(\ell_1^2+\ell_2^2));
\eqno(3.96)$$
$$\iota_1=m\es_1\sqrt{\frac{2(b_1c_2-b_2c_1)(\ell_1^2+\ell_2^2)}{a_1b_2-a_2b_1}},\;\;
\iota_2=m\es_2\sqrt{\frac{2(a_2c_1-a_1c_2)(\ell_1^2+\ell_2^2)}{a_1b_2-a_2b_1}},
,\eqno(3.97)$$ $\Im=\sn\varpi$ and
$(k_1,k_2)=-(1+m^2)(\ell_1^2+\ell_2^2)(c_1,c_2);$
$$\iota_1=m\es_1\sqrt{\frac{2(b_2c_1-b_1c_2)(\ell_1^2+\ell_2^2)}{a_1b_2-a_2b_1}},
\;\;
\iota_2=m\es_2\sqrt{\frac{2(a_1c_2-a_2c_1)(\ell_1^2+\ell_2^2)}{a_1b_2-a_2b_1}},
\eqno(3.98)$$ $\Im=\cn\varpi$ and
$(k_1,k_2)=(2m^2-1)(\ell_1^2+\ell_2^2)(c_1,c_2);$
$$\iota_1=\es_1\sqrt{\frac{2(b_2c_1-b_1c_2)(\ell_1^2+\ell_2^2)}{a_1b_2-a_2b_1}},\;
\;
\iota_2=\es_2\sqrt{\frac{2(a_1c_2-a_2c_1)(\ell_1^2+\ell_2^2)}{a_1b_2-a_2b_1}},
\eqno(3.99)$$ $\Im=\dn\varpi$ and
$(k_1,k_2)=(2-m^2)(\ell_1^2+\ell_2^2)(c_1,c_2).$ \psp

{\bf Theorem 3.7}. {\it Let $d_1,d_2,\ell_1,\ell_2,\ell_3\in\mbb{R}$
and let $\es_1,\es_2\in\{1,-1\}$. We have the following solutions of
the coupled two-dimensional cubic nonlinear Schr\"{o}dinger
equations (1.2) and (1.3)}:
$$\psi=\frac{\es_1e^{(x^2+y^2)i/4c_1t}}{\ell_1x+\ell_2y+\ell_3t}\sqrt{\frac{2(b_1c_2-b_2c_1)(\ell_1^2+\ell_2^2)}
{a_1b_2-a_2b_1}},\eqno(3.100)$$
$$\vf=\frac{\es_2e^{(x^2+y^2)i/4c_2t}}{\ell_1x+\ell_2y+\ell_3t}
\sqrt{\frac{2(a_2c_1-a_1c_2)(\ell_1^2+\ell_2^2)}
{a_1b_2-a_2b_1}};\eqno(3.101)$$
$$\psi=\es_1e^{(x^2+y^2)i/4c_1t}\sqrt{\frac{b_1c_2-b_2c_1}{(a_1b_2-a_2b_1)
((x-d_1t)^2+(y-d_2t)^2)}}, \eqno(3.102)$$
$$\vf=\es_2e^{(x^2+y^2)i/4c_2t}
\sqrt{\frac{a_2c_1-a_1c_2}{(a_1b_2-a_2b_1)((x-d_1t)^2+(y-d_2t)^2)}};
\eqno(3.103)$$
\begin{eqnarray*}\hspace{2cm}\psi&=&
\es_1\sqrt{\frac{2(b_1c_2-b_2c_1)
(\ell_1^2+\ell_2^2)}{a_1b_2-a_2b_1}}\;\frac{1}{t}\tan\frac{\ell_1x+\ell_2y+\ell_3t}{t}\\
& &\times\exp\left(\frac{x^2+y^2}{4c_1t}
-\frac{2c_1(\ell_1^2+\ell_2^2)}{t}\right)i,\hspace{5.3cm}(3.104)\end{eqnarray*}
\begin{eqnarray*}\hspace{2cm}\vf&=&\es_2
\sqrt{\frac{2(a_2c_1-a_1c_2)(\ell_1^2+\ell_2^2)}{a_1b_2-a_2b_1}}\;\frac{1}{t}
\tan\frac{\ell_1x+\ell_2y+\ell_3t}{t}\\
& &\times\exp\left(\frac{x^2+y^2}{4c_2t}
-\frac{2c_2(\ell_1^2+\ell_2^2)}{t}\right)i;\hspace{5.3cm}(3.105)\end{eqnarray*}
\begin{eqnarray*}\hspace{2cm}\psi&=&\es_1
\sqrt{\frac{2(b_1c_2-b_2c_1)
(\ell_1^2+\ell_2^2)}{a_1b_2-a_2b_1}}\;\frac{1}{t}\sec\frac{\ell_1x+\ell_2y+\ell_3t}{t}\\
& &\times\exp\left(\frac{x^2+y^2}{4c_1t}
+\frac{c_1(\ell_1^2+\ell_2^2)}{t}\right)i,\hspace{5.5cm}(3.106)\end{eqnarray*}
\begin{eqnarray*}\hspace{2cm}\vf&=&\es_2
\sqrt{\frac{2(a_2c_1-a_1c_2)(\ell_1^2+\ell_2^2)}{a_1b_2-a_2b_1}}\;\frac{1}{t}
\sec\frac{\ell_1x+\ell_2y+\ell_3t}{t}\\
& &\times\exp\left(\frac{x^2+y^2}{4c_2t}
+\frac{c_2(\ell_1^2+\ell_2^2)}{t}\right)i;\hspace{5.5cm}(3.107)\end{eqnarray*}
\begin{eqnarray*}\hspace{2cm}\psi&=&\es_1
\sqrt{\frac{2(b_1c_2-b_2c_1)
(\ell_1^2+\ell_2^2)}{a_1b_2-a_2b_1}}\;\frac{1}{t}\coth\frac{\ell_1x+\ell_2y+\ell_3t}{t}\\
& &\times\exp\left(\frac{x^2+y^2}{4c_1t}
+\frac{2c_1(\ell_1^2+\ell_2^2)}{t}\right)i,\hspace{5.3cm}(3.108)\end{eqnarray*}
\begin{eqnarray*}\hspace{2cm}\vf&=&\es_2
\sqrt{\frac{2(a_2c_1-a_1c_2)(\ell_1^2+\ell_2^2)}{a_1b_2-a_2b_1}}\;\frac{1}{t}
\coth\frac{\ell_1x+\ell_2y+\ell_3t}{t}\\
& &\times\exp\left(\frac{x^2+y^2}{4c_2t}
+\frac{2c_2(\ell_1^2+\ell_2^2)}{t}\right)i;\hspace{5.3cm}(3.109)\end{eqnarray*}
\begin{eqnarray*}\hspace{2cm}\psi&=&\es_1
\sqrt{\frac{2(b_1c_2-b_2c_1)
(\ell_1^2+\ell_2^2)}{a_1b_2-a_2b_1}}\;\frac{1}
{t}\csch\frac{\ell_1x+\ell_2y+\ell_3t}{t}\\
& &\times\exp\left(\frac{x^2+y^2}{4c_1t}
-\frac{c_1(\ell_1^2+\ell_2^2)}{t}\right)i,\hspace{5.5cm}(3.110)\end{eqnarray*}
\begin{eqnarray*}\hspace{2cm}\vf&=&\es_2
\sqrt{\frac{2(a_2c_1-a_1c_2)(\ell_1^2+\ell_2^2)}{a_1b_2-a_2b_1}}\;
\frac{1}{t}\csch\frac{\ell_1x+\ell_2y+\ell_3t}{t}\\
& &\times\exp\left(\frac{x^2+y^2}{4c_2t}
-\frac{c_2(\ell_1^2+\ell_2^2)}{t}\right)i;\hspace{5.5cm}(3.111)\end{eqnarray*}
\begin{eqnarray*}\hspace{2cm}\psi&=&
m\es_1\sqrt{\frac{2(b_1c_2-b_2c_1)
(\ell_1^2+\ell_2^2)}{a_1b_2-a_2b_1}}\;\frac{1}{t}\sn\frac{\ell_1x+\ell_2y+\ell_3t}{t}\\
& &\times\exp\left(\frac{x^2+y^2}{4c_1t}
+\frac{c_1(1+m^2)(\ell_1^2+\ell_2^2)}{t}\right)i,\hspace{3.9cm}(3.112)\end{eqnarray*}
\begin{eqnarray*}\hspace{2cm}\vf&=&
m\es_2\sqrt{\frac{2(a_2c_1-a_1c_2)(\ell_1^2+\ell_2^2)}{a_1b_2-a_2b_1}}\;\frac{1}{t}
\sn\frac{\ell_1x+\ell_2y+\ell_3t}{t}\\
& &\times\exp\left(\frac{x^2+y^2}{4c_2t}
+\frac{c_2(1+m^2)(\ell_1^2+\ell_2^2)}{t}\right)i;\hspace{3.9cm}(3.113)\end{eqnarray*}
\begin{eqnarray*}\hspace{2cm}\psi&=&
m\es_1\sqrt{\frac{2(b_2c_1-b_1c_2)(\ell_1^2+\ell_2^2)}{a_1b_2-a_2b_1}}\;\frac{1}{t}\cn\frac{\ell_1x+\ell_2y+\ell_3t}{t}\\
& &\times\exp\left(\frac{x^2+y^2}{4c_1t}
-\frac{c_1(2m^2-1)(\ell_1^2+\ell_2^2)}{t}\right)i,\hspace{3.7cm}(3.114)\end{eqnarray*}
\begin{eqnarray*}\hspace{2cm}\vf&=&
m\es_2\sqrt{\frac{2(a_1c_2-a_2c_1)(\ell_1^2+\ell_2^2)}{a_1b_2-a_2b_1}}\;\frac{1}{t}
\cn\frac{\ell_1x+\ell_2y+\ell_3t}{t}\\
& &\times\exp\left(\frac{x^2+y^2}{4c_2t}
-\frac{c_2(2m^2-1)(\ell_1^2+\ell_2^2)}{t}\right)i;\hspace{3.6cm}
(3.115)\end{eqnarray*}
\begin{eqnarray*}\hspace{2.2cm}\psi&=&\es_1
\sqrt{\frac{2(b_2c_1-b_1c_2)(\ell_1^2+\ell_2^2)}{a_1b_2-a_2b_1}}\;\frac{1}{t}\dn\frac{\ell_1x+\ell_2y+\ell_3t}{t}\\
& &\times\exp\left(\frac{x^2+y^2}{4c_1t}
+\frac{c_1(m^2-2)(\ell_1^2+\ell_2^2)}{t}\right)i,\hspace{3.7cm}(3.116)\end{eqnarray*}
\begin{eqnarray*}\hspace{2.2cm}\vf&=&\es_2
\sqrt{\frac{2(a_1c_2-a_2c_1)(\ell_1^2+\ell_2^2)}{a_1b_2-a_2b_1}}\;\frac{1}{t}
\dn\frac{\ell_1x+\ell_2y+\ell_3t}{t}\\
& &\times\exp\left(\frac{x^2+y^2}{4c_2t}
+\frac{c_2(m^2-2)(\ell_1^2+\ell_2^2)}{t}\right)i.\hspace{3.8cm}
(3.117)\end{eqnarray*}

 \vspace{0.5cm}

\noindent{\Large \bf References} \vspace{0.2cm}

\begin{description}

\item[{[AEK]}] N. Akhmediev, V. Eleonskii and N. Kulagin,
First-order exact solutions of the nonlinear Schr\"{o}dinger
equation, {\it Teoret. Mat. Fiz.} {\bf 72} (1987), 183-196.

\item[{[GG]}] B. Gr\'{e}bert and J. Guillot, Periodic solutions of
coupled nonlinear Schr\"{o}dinger equations in nonlinear optics:
the resonant case, {\it  Appl. Math. Lett.} {\bf 9} (1996), 65-68.

\item[{[GW]}] L. Gagnon and P. Winternitz, Exact solutions of the
cubic and quintic nonlinear Schr\"{o}dinger equation for a
cylindrical geometry, {\it Phys. Rev. A} {\bf 22} (1989), 296

\item[{[HS]}] F. Hioe and T. Salter, Special set and solutions of
coupled nonlinear Schr\"{o}dinger equations, {\it J. Phys. A:
Math. Gen.} {\bf 35} (2002), no. 42, 8913-8928.

\item[{[I]}] N. H. Ibragimov, {\it Lie Group Analysis of
Differential Equations}, Volume 2, CRC Handbook, CRC Press, 1995.

\item[{[MP]}] D. Mihalache and N. Panoin, Exact solutions of
 nonlinear Schr\"{o}dinger equation for positive group velocity
 dispersion, {\it J. Math. Phys.} {\bf 33} (1992), no. 6,
 2323-2328.

\item[{[RL1]}] R. Radhakrishnan and M. Lakshmanan, Bright and dark
soliton solutions to coupled nonlinear Schr\"{o}dinger equations,
{\it J. Phys. A: Math. Gen.} {\bf 28} (1995), no. 9, 2683-2692.

\item[{[RL2]}] R. Radhakrishnan and M. Lakshmanan, Exact soliton
solutions to coupled nonlinear Schr\"{o}dinger equations with
higher-order effects, {\it Phys. Rev. E(3)} {\bf 54} (1995), no.
3, 2949-2955.

\item[{[SEG]}] E. Saied, R. EI-Rahman and M. Ghonamy, On the exact
 solution of (2+1)-dimensional cubic nonlinear Schr\"{o}dinger
(NLS) equation", {\it J. Phy. A: Math. Gen.} {\bf 36} (2003),
6751-6770.

\item[{[WG]}] Z. Wang and D. Guo, {\it Special functions}, World
Scientific, Singapore, 1998.

\end{description}

\end{document}